\documentclass [superscriptaddress,preprint,amsmath,amssymb,showkeys,floatfix]{revtex4-2}
\usepackage{xcolor}
\usepackage{soul}
\usepackage[normalem]{ulem}
\usepackage[hidelinks]{hyperref}
\usepackage{graphicx}
\usepackage{dcolumn}
\usepackage{bm}
\usepackage{color}
\usepackage{float}
\usepackage[utf8]{inputenc}
\usepackage[mathlines]{lineno}
\usepackage{multirow}
\usepackage{amsmath}
\usepackage{hyperref}
\usepackage{soul,xcolor}
\usepackage{natbib}
\usepackage{amsmath} 
\usepackage{pdfpages}
\usepackage{pgffor}
\usepackage{xcolor}
\usepackage{xr}
\usepackage{chemformula}
\usepackage{siunitx}
\usepackage{physics}
\soulregister\cite7
\soulregister\ref7
\setstcolor{blue}
\newcommand{\doHMN}[2]{%
  \begingroup\lccode`~=`#1
  \lowercase{\endgroup\let~}#2%
  \mathcode`#1="8000
}

\makeatletter
\AtBeginDocument{\let\LS@rot\@undefined}
\newcommand*{\addFileDependency}[1]{
  \typeout{(#1)}
  \@addtofilelist{#1}
  \IfFileExists{#1}{}{\typeout{No file #1.}}
}
\makeatother

\usepackage{tikz}

\begin{document}
\title{Robust Control over Polar Skyrmion Bubble Density with a Combined Optical and Electrical Approach}
\author{Lingyuan Gao}
\thanks{Corresponding author: lg041@uark.edu}
\affiliation{Smart Ferroic Materials Center, Physics Department and Institute for Nanoscience and Engineering, University of Arkansas, Fayetteville, Arkansas, 72701, USA}
\author{Laurent Bellaiche}
\thanks{Corresponding author: laurent@uark.edu}
\affiliation{Smart Ferroic Materials Center, Physics Department and Institute for Nanoscience and Engineering, University of Arkansas, Fayetteville, Arkansas, 72701, USA}
\affiliation{Department of Materials Science and Engineering, Tel Aviv University, Ramat Aviv, Tel Aviv 6997801, Israel}

\begin{abstract}
Polar skyrmion bubbles are nanoscale ferroelectric domain configurations with swirling polarization textures, and often emerge in ferroelectric oxide systems. Owing to their inhomogeneous polarization patterns, which endow them with distinct topologies and electrical responses from homogeneous monodomains,  polar skyrmion bubbles are envisaged to be promising candidates for non-volatile memory devices. In such device, the recorded information density is directly proportional to the density of bubbles, underscoring the need for precise control over bubble nucleation. Here, using first-principles-based calculations, we demonstrate that when assisted with a DC electric field, twisted light, which has a spatially inhomogeneous field pattern, can robustly tune the density of polar skyrmion bubbles in ferroelectric ultrathin films between $10^2\sim 10^4 \rm{bit}/\mu m^2$. Moreover, by modulating DC and optical field strengths together with the beam radius, the nucleation rate, which characterizes the creation and annihilation speed of polar skyrmion bubbles, can also be well controlled. These findings highlight the unique response of ferroelectric nanofilms to optical and electric fields, which is crucial for employing polar skyrmion bubbles in the next-generation of ultrahigh-density memory technologies.  
 
\end{abstract}

\maketitle
\titlepage
\section*{Introduction}
Complex interplay between elastic, electrostatic, and gradient energies in ferroelectric systems give rise to diverse structures formed by electric dipoles~\cite{chen2021recent,junquera2023topological,govinden2023spherical,wang2023entangled}. Among them, polar skyrmion bubbles, which are often observed in ferroelectric oxide superlattices, are drawing great attention~\cite{nahas2015discovery,zhang2017nanoscale,das2019observation, pereira2019theoretical}. With in-plane and out-of-plane polarization components rotating along azimuthal and radial directions respectively on the two-dimensional plane, the structure is endowed with a topological number ``1", setting it distinct from other configurations~\cite{nagaosa2013topological}. This does not only guarantee its stability against perturbations but also present skyrmion bubble as a potential ``bit" to encode and carry information~\cite{fert2013skyrmions,tomasello2014strategy,luo2021skyrmion}. 

One critical question regarding the application of polar skyrmion bubbles is how to effectively write skyrmions into ferroelectric systems. If the number of bubbles can be altered at will, the information storage density will be robustly manipulated, which is a crucial step towards realizing ultrahigh-density memory devices~\cite{han2022high,xue2025observation}. Previous studies show that in the design of the ferroelectric superlattices $((\rm{PbTiO_3})_n/(SrTiO_3)_n)_m$ (PTO/STO), bubble density increases with the periodicity $m$~\cite{das2019observation}, but decreases with the $\rm PbTiO_3$ layer thickness $n$~\cite{gong2023absence}. In addition, owing to the skyrmion formation origin, electric field and mechanical strain have also been proposed as reliable strategies for creating and annihilating polar bubbles: \textit{ab initio} simulations reveal that quasihexagonal arrays of electric bubbles form in ferroelectric PbZr$_{0.4}$Ti$_{0.6}$O$_3$ ultrathin films
under external electric field at low temperatures~\cite{makeev2024quasihexagonal}; phase-field simulations reveal that by applying electrodes, the DC electric field can reduce the total number of skyrmion bubbles in PTO/STO superlattice by two-thirds in a nonvolatile manner~\cite{zhou2022local}; an \emph{in situ} scanning
transmission electron microscopy (STEM) observation further confirms that electric field can increase the number of bubbles in the same system by fragmenting original skyrmions into smaller skyrmions, and the field along the opposite direction can expand skyrmion size and drive their merging~\cite{zhu2022dynamics}. As to the mechanical stimuli, one phase-field study shows that a compressive strain can lead to a reduction of skyrmion number also in PTO/STO superlattice~\cite{zhang2022strain}, while vibrational tapping with scanning probe microscope tips can accurately write polar skyrmion bubbles in BiFeO$_3$ thin films~\cite{kim2025dynamic}; in van der Waals ferroelectric crystal CuInP$_2$S$_6$, mechanical force can tailor the coexistence of  polar phases, enabling the formation of high-density bubble states~\cite{jiang2025mechanically}. A very recent computational study also shows that by twisting two monolayers of paraelectric oxide SrTiO$_3$, high-density polar skyrmion latttice with bubble size at 1 nanometer (nm) can be established, which results from the  strong interlayer coupling and the unique strain fields in Moir\'e structure~\cite{xu2025creating}. In the magnetic counterpart, beyond the electric field/voltage, and magnetic field~\cite{schott2017skyrmion,wang2018ferroelectrically,denker2023size,bhattacharya2020creation,lone2024controlling,chen2024all}, the control over magnetic skyrmion density can also be achieved via current pulses~\cite{ang2020electrical}, ion irradiation~\cite{hu2022precise}, and ultrafast laser pulse~\cite{kern2022tailoring,je2018creation}. A recent theoretical study developed an effective model for laser-induced skyrmion nucleation, predicting that the stabilized skyrmion number depends on the cooling time following the incident heat pulse introduced by the laser~\cite{liefferink2025effective}. These studies inspire us to study whether we can employ analogous dynamical approaches on polar skyrmions.   

Molecular dynamics simulations in our previous study showed that how an optical vortex beam, also known as twisted light, can induce a single, dynamically evolving skyrmion in the irradiated region of Pb(Zr$_{0.4}$Ti$_{0.6}$)O$_3$ (PZT) ferroelectric ultrathin films~\cite{gao2024dynamical}. However, under illumination, no additional bubbles emerge elsewhere in the system. This limitation is disadvantageous for the design of ultrahigh density memristors, which requires the existence of multiple bubbles to function as bits. Therefore, it is imperative to explore whether ferroelectric systems can sustain multiple bubbles under light, which would pave the way for memory devices operated through optical control. 

In this work, we computationally demonstrate that twisted light can induce multiple polar skyrmion bubbles, rather than a single skyrmion, out of a homogeneous ferroelectric monodomain. A crucial ingredient here is a homogeneous DC electric field applied along the out-of-plane direction, along with the optical field: the twisted light introduces electrostatic instabilities beyond the illuminated region mediated by dipole-dipole interaction, while the DC electric field triggers nucleation and promotes the emergence of skyrmion bubbles. By varying the DC field magnitude, the number of polar skyrmion bubbles can be accurately and continuously controlled over a wide range. Moreover, by adjusting other optical field parameters, we find that the nucleation rate can be significantly slowed, with the relaxation process persisting over a hundred picoseconds (ps), ultimately arriving at a steady state with ultrahigh bubble density. These results highlight the synergy between the electrical and optical approaches in driving exotic phase transitions, enabling the condensation of polar skyrmion bubbles and polarization switching.

\section*{RESULTS}
We model the interaction between twisted light and PZT ferroelectric ultrathin films using the same first-principles-based effective Hamiltonian molecular dynamics framework as in previous studies~\cite{zhong1994phase,zhong1995first,gao2024dynamical}. The simulation cell is chosen as a $80\times 80 \times 5$ unit cell (u.c.), corresponding to a thickness of 5 layers and an in-plane dimension about 40 nm with periodic boundary conditions imposed along in-plane axes. The initial configuration is established as a down-poled monodomain along the out-of-plane [00$\bar{1}$] direction thermalized at 10 K. More computational details are provided in Methods section. Note that neither the number of layers (even or odd) nor the initial monodomain polarization direction (upward or downward) alter the induced phenomena, and the corresponding calculation results are provided in the supplementary information (SI).

The twisted light, which is the lowest order Laguerre-Gaussian beam ($p = 0, l = 1$) with left-handed circular polarization~\cite{allen1992orbital}, is introduced to constantly illuminate the PZT film. The light is normally incident on the center of the $xy$ plane, so that its electric field is coupled with in-plane dipole components $p_{x,y}$. In addition, an upward DC electric field is also applied along the out-of-plane $+z$ direction. In our calculations, we use a light with a representative $f = 1$ THz frequency and a beam radius of 5 u.c. for driving motions of ions. Discussions on chosen light parameters are referred to the Methods section. In contrast to our previous work, where a single skyrmion directly induced by twisted light on surface layers is robust over a broad frequency range of 0.2$\sim$2.5 THz~\cite{gao2024dynamical}, here multiple skyrmion bubbles that form on subsurface layers survive within a narrow frequency window of $0.97\sim 1.07$ THz, and the relevant discussion on frequency dependence is presented in SI. Below we use $E_0$ and $E_{\rm{DC}}$ to denote the magnitudes of the twisted light and the DC electric field, respectively. The interaction process is illustrated in Fig. 1.   

\subsection*{Robust formation of polar skyrmion bubbles}
In a previous work~\cite{gao2024dynamical}, the dynamical polar skyrmions on surface layers are activated when $E_{0}$ varies between $100\sim 300$ MV/cm. Here, we first fix $E_{0}$ at a medium value $180$ MV/cm and $E_{\rm{DC}} = 0.725$ MV/cm. $N_{\rm{sk}}$ is computed upon the unit vector of local electric dipole $\bm{n}(\vec{r},t)$ on the layer~\cite{nagaosa2013topological}:
\begin{equation}
    N_{\rm{sk}} = \int \rho_{\rm{sk}}(\vec{r}) d^2\vec{r} = \int \frac{1}{4\pi}\bm{n}(\vec{r},t)\cdot\big( \frac{\partial \bm{n}(\vec{r},t)}{\partial x} \times \frac{\partial \bm{n}(\vec{r},t)}{\partial y} \big) d^2 \vec{r}.
\end{equation}
On discrete lattices, $\rho_{\rm{sk}}(\vec{r})$ is computed as the area of spherical triangle in the order parameter space~\cite{berg1981definition}. As shown in Fig. 2(a), $N_{\rm{sk}}$ at the top and bottom surface layers alternate between ``0" and ``1", signaling a continuous dynamical evolution of polar skyrmions induced by the twisted light, consistent with the previous work~\cite{gao2024dynamical}. However, by applying the DC field, $N_{\rm{sk}}$ on layer 2 and layer 4 (subsurface layers adjacent to the surface layers) deviates from zero, in contrast to the behavior observed without the DC field. $N_{\rm{sk}}$ rises in the first 2 ps and then stabilizes at a value of 18. The dipole patterns in layer 2 and layer 4 are visualized in Fig. 2(b). At the center of the plane, dipoles form a vortex with notable swirling $p_{x,y}$ due to coupling with the twisted light field. whereas the out-of-plane $p_z$ components there are negative and parallel to the monodomain background, indicating that no skyrmion forms there. Conversely, 18 skyrmion bubbles are distributed inhomogeneously across the $xy$ plane, away from the irradiated region. They reside at identical in-plane sites in the two layers, representing cross sections of flux-closure electric bubbles at different latitudes~\cite{prokhorenko2024motion}. A 3D visualization of a skyrmion bubble is given in Fig. 2(c). These skyrmions have relatively smaller sizes compared to the central vortex, and they are mostly ``N\'eel" type, with opposite helicities on the two layers. With time evolution of the twisted light field, in-plane components of the central vortex drastically changes over time (e.g., from an approximately divergent to counterclowise pattern, as shown in Fig. 2(b)); nevertheless, the ``N\'eel"-type character and the positions of 18 skyrmion bubbles remain stable, and they do not change with the vortex texture. 

We then examine the effect of the DC field magnitude. The results are presented in Fig. 3. When $E_{\rm{DC}} \leq 0.64$ MV/cm, no skyrmion bubbles on the subsurface layers are activated, although the signature of central vortex is pronounced. Above this critical value, skyrmion bubbles start to nucleate: when $E_{\rm{DC}} = 0.65$ MV/cm, $N_{\rm{sk}}$ increases to 3, evidenced by three small skyrmions nucleated in the regions far from the center, as presented in Fig. 3(b). Between 0.65$\sim$0.69 MV/cm, $N_{\rm{sk}}$ does not change much with $E_{\rm{DC}}$ until when it reaches 0.7 MV/cm, where 10 skyrmion bubbles are nucleated. After that, $N_{\rm{sk}}$ rises sharply with $E_{\rm{DC}}$, and it reaches its maximum value 18 when $E_{\rm{DC}}$ = 0.725 MV/cm. A further increase on $E_{\rm{DC}}$ to 0.73 MV/cm annihilates all bubbles, which will be discussed in details later.

\subsection*{Polar skyrmion bubble formation as a nonlocal effect}
With the exponential decay of the twisted light magnitude, the optical field at bubble positions has minor effects. This demonstrates that the skyrmion formation is a nonlocal process, rather than originating from the coupling between the twisted light and electric dipoles, which is responsible for the dynamical skyrmion formed on surface layers~\cite{gao2024dynamical}. Nevertheless, mediated by long-range dipole-dipole interaction, which is critical in ferroelectric system~\cite{zhong1995first,ponomareva2005low}, the variation of dipoles directly governed by twisted light in the central region can extend to the regions far from the center, and modify dipole configurations there. Owing to the nature of dipole response that acts to neutralize the system and lower the total electrostatic energy~\cite{luk2020hopfions,tikhonov2022polarization}, variations of $p_{x,y}$ components can impose changes on $p_z$, and even flip its orientation~\cite{gao2024dynamical}. An added upward DC field along $+z$ direction can further assist the flipping process, and introduce additional electrostatic instability to the system, contributing to the nonlocal nucleation of polar skyrmion bubbles. Note that here, the optical-electrical writing of bubbles in a monodomain is a volatile process: once the light is removed, the bubbles recover to the poled-down monodomain. However, previous studies have shown that non-volatile writing is possible when the initial state is a stripe domain, where the energy barriers between metastable states are lower than in the monodomain case~\cite{gao2024dynamical,gao2025poincare}.

With the bubble nucleation being nonlocal to the optical field, it is intriguing to explore whether it is a random or a deterministic process. To address this question, we perform three independent simulations with different initial thermalizations. As we find across all three simulations, all three skyrmion bubbles shown in Fig. 3(b) emerge at identical locations despite differences in the initial monodomain configurations, indicating that until this step, the nucleation is deterministic. To understand the bubble formation mechanism, we plot the corresponding spatial distribution of long-range dipole-dipole interaction energy on the subsurface layer in Fig. 3(c), and we do see that the locations where bubbles reside correspond to slightly lower interaction energies. The deterministic selection of these specific nucleation sites is not fully understood, since  they are not distinguishable from other sites energy-wise at the beginning of the simulation. We infer it may originate from subtle, nonlocal correlations mediated by dipole–dipole interactions during the dipole evolution. However, when eleven bubbles are stablized at $E_{DC}\approx 0.71$ MV/cm, four bubbles appear at different locations across three simulations, as shown by Figs. 3(d)-(f), indicating that when more bubbles are present, their nucleation is more random and is affected by thermal fluctuations. An additional note is that at higher temperatures, the bubble formation is highly sensitive to thermal fluctuation; even at 30 K, which is still relatively low, the system is only marginally able to sustain bubbles (see detailed discussion in SI).

\subsection*{Annihilation of polar skyrmion bubbles}
We next focus on $E_{\rm{DC}}= 0.73$ MV/cm, when all skyrmions are annihilated eventually. Figure 4(a) shows the variation of $N_{\rm{sk}}$ with time on each layer; on two subsurface layers, $N_{\rm{sk}}$ gradually increases and reaches its positive maximum at $\sim60$, then decreases and switches to a negative value, and continues to grow toward a negative maximum $\sim-70$; subsequently, $|N_{\rm{sk}}|$ decreases and approaches zero around 5.5 ps. The dipole evolution on subsurface layer 4 at different stages are shown in Fig. 4(b). After $N_{\rm{sk}}$ reaches 60 at 3.9 ps, neighboring bubbles begin to connect and merge, leading to an expansion of individual bubble size and a decrease of $N_{\rm{sk}}$, as illustrated for 4.4 ps; at one point, such connection turns the majority connected polarization background from downward to upward, causing $N_{\rm{sk}}$ to switch from being positive to negative at 4.5 ps; at 4.9 ps, $N_{\rm{sk}}$ reaches -74, with the core of each bubble exhibiting downward polarization; the core area shrinks with time, and finally the system arrives at an up-poled monodomain with a central vortex. For two surface layers 1 and 5, they exhibit similar polarization switching from downward to upward with the intermediate bubble creation and annihilation, but on a faster timescale from 3.5 to 4.5 ps. Eventually $N_{\rm{sk}}$ on these two surface layers alternate between ``0" and ``1"  as a response to the applied twisted light. Therefore, in the presence of both dynamical optical field and static electric field, the out-of-plane polarization proceeds through a nontrivial pathway, along which multiple bubbles emerge as stable, intermediate states.

Another crucial factor is the beam radius of the twisted light, as that determines areas of dipoles directly coupled with the light. Considering the dimension of the film, we double the beam radius to 10 u.c., allowing fields with appreciable magnitude to extend over a larger region. Without presenting the full spectrum of $E_{0}$ and $E_{\rm{DC}}$ in this case, we highlight one intriguing scenario when $E_{\rm{DC}}$ = 0.3 MV/cm and $E_{0}$ ranging between 50 and 90 MV/cm, where a slow skyrmion nucleation process is observed. Figure 5(a) shows the variation of $N_{\rm{sk}}$ when $E_{0}$=80 MV/cm. For both surface layers and subsurface layers, the respective $N_{\rm{sk}}$ exhibits similar increasing and decreasing behavior as in Fig. 4(a), but persists over 100 ps. Notably, $N_{\rm{sk}}$ on two subsurface layers are stabilized after 130 ps, indicating that the system reaches a steady state. Such steady state with ultrahigh density skyrmion bubbles cannot be maintained with a smaller beam radius, and will relax to a up-poled monodomain when optical and electric fields are removed. With relaxation occurring at a timescale of 100 ps, this dynamical process should be more readily captured by time-resolved experimental techniques~\cite{wang2024giant,li2025terahertz,wang2025terahertz}. In addition, we investigate the impact of the simulation cell size. Figure 5(b) shows the variation of stabilized $N_{\rm{sk}}$ on the subsurface layer with the DC field magnitude in a $100 \times 100 \times 5$ simulation cell when $E_0 = 180$ MV/cm. Compared to Fig. 3(a), the tuning of $N_{\rm{sk}}$ by the DC field is also robust in the larger cell, but the maximal bubble density is 1.2 times larger, which can be attributed to a larger sample-to-beam size ratio that can accommodate more bubbles; in the meantime, a higher DC field magnitude is required to stabilize them.

\subsection*{Tunability of polar skyrmion bubbles}
We finally turn to investigate whether the control over bubble density is robust by varying field magnitudes and beam morphologies. To the first end, we vary $E_{0}$ from 90 MV/cm up to 300 MV/cm in steps of 60 MV/cm, and for each case we scan $E_{\rm{DC}}$ to seek emergent skyrmion bubbles. The results are summarized as a histogram shown in Fig. 5(c). The variations of $N_{\rm{sk}}$ with $E_{\rm{DC}}$ at other $E_0$s are very similar to the behavior at $E_{0}$= 180 MV/cm. A general trend is that at larger $E_{0}$, the ferroelectric system sustain fewer skyrmions---up to 14 skyrmion bubbles emerge when $E_{0}$= 300 MV/cm. Also, the window of $E_{\rm{DC}}$ to support active bubbles is wider at larger $E_{0}$. For example, at $E_{0}$= 90 MV/cm, $E_{\rm{DC}}$ allowing for skyrmion nucleation ranges in 0.82$\sim$0.86 MV/cm , and the window is broadened to 0.6$\sim$0.7 MV/cm for $E_{0}$= 300 MV/cm.
To the second end, one may naturally wonder whether a purely circularly polarized light, which does not carry orbital angular momentum and is spatially homogeneous, can exert a similar effect in yielding polar bubbles. We thus adopt a Gaussian profile for a left-handed circular-polarized (CP) light pulse with 5 u.c. as the beam radius and 180 MV/cm as the field magnitude. As shown in the inset of Fig. 4(b), the field is homogeneously aligned within the beam radius but changes its orientation with time. Figure 4(b) shows that up to 7 polar skyrmion bubbles can be induced by the Gaussian pulse, less than one half of $N_{\rm{sk}}$ generated under twisted light of the same magnitude. The active $E_{\rm{DC}}$ window is also narrower than that under twisted light as shown in Fig. 3(a). This indicates that the spatial inhomogeneity of the optical field pattern is important in giving rise to more polar skyrmion bubbles.

\section*{Discussion}
In summary, using first-principles-based techniques, we demonstrate that a combined approach that utilizes twisted light and DC electric field can induce multiple polar skyrmion bubbles in ferroelectric ultrathin films, which are intermediate states during polarization switching. The stabilization of the intermediate multi-bubble states results from a concerted effort between local optical field and global DC electric field, and the former affects the nonlocal bubble nucleation through a long-range dipole-dipole interaction. A key advantage of this strategy is, by tuning field parameters and the interplay between optical and electric fields, bubble density can be robustly controlled. The results are robust regarding to magnitudes of both AC and DC electric fields and the variation of beam radius but are sensitive to temperature and optical field frequency. In particular, we reveal an ultraslow skyrmion nucleation process on the order of hundred picoseconds, which eventually transitions to a steady state with ultrahigh density bubbles, more easily accessible to experimental observation. Our work thus provides a viable pathway toward high-density memory devices based on polar skyrmions.

\section*{METHODS}
\subsection*{Effective-Hamiltonian Molecular Dynamics Simulations}

We model ferroelectric ultrathin PZT films grown along the pseudo-cubic [001] axis, indicating that the spontaneous polarization is along the out-of-plane direction; nevertheless, the termination of polarization at the surface induces  bound charges there, which generate a strong depolarization field that opposes the polarization and leads to the development of in-plane polarization components; furthermore, owing to the interplay between electrostatic energy, elastic energy and their mutual coupling, polarization often forms swirling textures such as vortex and skyrmion~\cite{chen2021recent,junquera2023topological}. We apply a $\sim$-2$\%$ compressive strain along the periodic [100] and [010] pseudo-cubic directions. Along the out-of-plane direction, the system is intentionally confined with one substrate layer and two vacuum layers attached to the bottom and top  of five PZT layers to mimic the depolarization field at the surface. We use a parameter $\beta=0.86$ to model the dielectric screening condition of the thin film, with $\beta =1$ and $\beta = 0$ denoting the full- and no- screening limits, respectively~\cite{ponomareva2005low}. In practice, to allow the interaction between light and ferroelectric films, graphene can be a potential candidate for top electrode with its conductiveness and excellent THz transmission~\cite{wu2013graphene}; in addition, since the optical beam is relatively small compared to the dimension of the ferroelectric monodomain, a ring-shape electrode with a central aperture as well as an integrated electrode by using wire metal grid pattern are suggested to allow high transmittance of THz light~\cite{wang2015broadband,fischer2023simultaneous}. We maintain constant pressure during the simulation (namely, an NPT ensemble), while relaxing strain and off-centering motions--two main degrees of freedoms (DOFs) associated with ferroelectric phase transitions. The former represents lattice deformation and accounts for elastic energy, whereas the latter is proportional to onsite electric dipoles, directly relevant to electrostatic energy. To model light-matter interaction, we add the Zeeman-like coupling term $-\sum_{i} \vec{E}_{i} \cdot \vec{p}_{i}$ into the total Hamiltonian~\cite{garcia1998electromechanical}, where $\vec{E}_i$ and $\vec{p}_{i}$ refer to the local electric field and dipole moment in cell $i$. Note that since thermal effects are also important in light illumination process, studying heat effects will be a worthy question to pursue in the future work. In addition, to further narrow the gap between calculation and experiments, it would be important to scale up the modeling dimensions to several hundred nanometers or even micrometers. 

\subsection*{Optical and Electrical Field Parameters}

We adopt a beam radius of 5 unit cells (u.c.) to ensure that the field decays to a negligible value when it arrives at the boundaries of the simulation cell; otherwise, there would be a field discontinuity at the boundary between neighboring supercells. With recent developments in plasmonics that can overcome the diffraction limit~\cite{prinz2023orbital,koya2023advances}, and that THz free-space light can be confined to  nanoscale~\cite{de2021nanoscale,kowalski2025ultraconfined,de2026enhanced}, these conditions hold promise for future feasibility. Also note that in the effective Hamiltonian work, field magnitudes are often scaled to be twenty times greater than experimental values~\cite{xu2017designing,jiang2018giant}. The dielectric breakdown DC field magnitude is between 0.3$\sim$1 MV/cm, well above the predicted $E_{\rm{DC}}$ values~\cite{ko2019improvement,nguyen2019experimental}. As to $E_0$, though the field maxima is above the dielectric breakdown limit, the field magnitude decays exponentially with the distance from the center. Furthermore, the dielectric constant of PZT film significantly drops at THZ frequency range~\cite{kwak2011dielectric}, indicating a much higher breakdown strength for a high-frequency optical field than for DC fields~\cite{mcpherson2002proposed}. Therefore, the required magnitudes of AC optical field and DC electric field are not expected to be impractically large and to cause damage to the system. However, we admit that focusing a THz twisted light to nanoscale is still a bottleneck for the current technology.

\section*{Declaration statements}
\noindent{\textbf{DATA AVAILABILITY}}
The datasets generated and/or analyzed during the current study are not publicly available due to the large size of the raw simulation trajectories and intermediate output files, but are available from the corresponding author on reasonable request.

\noindent{\textbf{CODE AVAILABILITY}}
All code and mathematical algorithm files are available from the corresponding author upon reasonable request.

\noindent{\textbf{ACKNOWLEDGMENTS}} We acknowledge the computational support from the High Performance Computing Modernization Program
of the Department of Defense and the Arkansas High Performance Computing Center. We acknowledge ARO Grants No. W911NF-21-2-0162 (ETHOS MURI) and No. W911NF-25-1-0223, and the support from
the Vannevar Bush Faculty Fellowship (VBFF) Grant No. N00014-20-1-2834 from the Department of Defense.

\noindent{\textbf{AUTHOR CONTRIBUTIONS}} L.G. conceived the idea, performed the molecular dynamics simulations, and wrote the manuscript. L.B. supervised the project. 

\noindent{\textbf{COMPETING INTERESTS}} The authors declare no competing interests.

\clearpage
\section*{references}
\bibliography{reference.bib}

@article{zhou2022local,
  title={Local manipulation and topological phase transitions of polar skyrmions},
  author={Zhou, Linming and Huang, Yuhui and Das, Sujit and Tang, Yunlong and Li, Cheng and Tian, He and Chen, Long-Qing and Wu, Yongjun and Ramesh, Ramamoorthy and Hong, Zijian},
  journal={Matter},
  volume={5},
  number={3},
  pages={1031--1041},
  year={2022},
  publisher={Elsevier}
}

@article{zhu2022dynamics,
  title={Dynamics of polar skyrmion bubbles under electric fields},
  author={Zhu, Ruixue and Jiang, Zhexin and Zhang, Xinxin and Zhong, Xiangli and Tan, Congbing and Liu, Mingwei and Sun, Yuanwei and Li, Xiaomei and Qi, Ruishi and Qu, Ke and others},
  journal={Physical Review Letters},
  volume={129},
  number={10},
  pages={107601},
  year={2022},
  publisher={APS}
}

@article{zhang2022strain,
  title={Strain manipulation of ferroelectric skyrmion bubbles in a freestanding {PbTiO$_3$} film: A phase field simulation},
  author={Zhang, Yixuan and Li, Qian and Huang, Houbing and Hong, Jiawang and Wang, Xueyun},
  journal={Physical Review B},
  volume={105},
  number={22},
  pages={224101},
  year={2022},
  publisher={APS}
}

@article{kim2025dynamic,
  title={Dynamic mechanical writing of skyrmion-like polar nanodomains},
  author={Kim, Jaegyu and Yeo, Youngki and Kwon, Yong-Jun and Lee, Juhyun and Seo, Jeongdae and Hong, Seungbum and Yang, Chan-Ho},
  journal={npj Quantum Materials},
  volume={10},
  number={1},
  pages={71},
  year={2025},
  publisher={Nature Publishing Group UK London}
}

@article{das2019observation,
  title={Observation of room-temperature polar skyrmions},
  author={Das, S and Tang, YL and Hong, Z and Gon{\c{c}}alves, MAP and McCarter, MR and Klewe, C and Nguyen, KX and G{\'o}mez-Ortiz, F and Shafer, P and Arenholz, E and others},
  journal={Nature},
  volume={568},
  number={7752},
  pages={368--372},
  year={2019},
  publisher={Nature Publishing Group UK London}
}

@article{gong2023absence,
  title={Absence of critical thickness for polar skyrmions with breaking the Kittel’s law},
  author={Gong, Feng-Hui and Tang, Yun-Long and Wang, Yu-Jia and Chen, Yu-Ting and Wu, Bo and Yang, Li-Xin and Zhu, Yin-Lian and Ma, Xiu-Liang},
  journal={Nature Communications},
  volume={14},
  number={1},
  pages={3376},
  year={2023},
  publisher={Nature Publishing Group UK London}
}

@article{de2021nanoscale,
  title={Nanoscale-confined terahertz polaritons in a van der Waals crystal},
  author={de Oliveira, Thales VAG and N{\"o}renberg, Tobias and {\'A}lvarez-P{\'e}rez, Gonzalo and Wehmeier, Lukas and Taboada-Guti{\'e}rrez, Javier and Obst, Maximilian and Hempel, Franz and Lee, Eduardo JH and Klopf, J Michael and Errea, Ion and others},
  journal={Advanced Materials},
  volume={33},
  number={2},
  pages={2005777},
  year={2021},
  publisher={Wiley Online Library}
}

@article{kowalski2025ultraconfined,
  title={Ultraconfined terahertz phonon polaritons in hafnium dichalcogenides},
  author={Kowalski, Ryan A and Mueller, Niclas S and {\'A}lvarez-P{\'e}rez, Gonzalo and Obst, Maximilian and Diaz-Granados, Katja and Carini, Giulia and Senarath, Aditha and Dixit, Saurabh and Niemann, Richarda and Iyer, Raghunandan B and others},
  journal={Nature Materials},
  volume={24},
  number={11},
  pages={1735--1741},
  year={2025},
  publisher={Nature Publishing Group UK London}
}

@article{gao2024dynamical,
  title={Dynamical control of topology in polar skyrmions via twisted light},
  author={Gao, Lingyuan and Prokhorenko, Sergei and Nahas, Yousra and Bellaiche, Laurent},
  journal={Physical Review Letters},
  volume={132},
  number={2},
  pages={026902},
  year={2024},
  publisher={APS}
}

@article{prokhorenko2024motion,
  title={Motion and teleportation of polar bubbles in low-dimensional ferroelectrics},
  author={Prokhorenko, S and Nahas, Y and Govinden, V and Zhang, Q and Valanoor, N and Bellaiche, L},
  journal={Nature Communications},
  volume={15},
  number={1},
  pages={412},
  year={2024},
  publisher={Nature Publishing Group UK London}
}

@article{junquera2023topological,
  title={Topological phases in polar oxide nanostructures},
  author={Junquera, Javier and Nahas, Yousra and Prokhorenko, Sergei and Bellaiche, Laurent and {\'I}{\~n}iguez, Jorge and Schlom, Darrell G and Chen, Long-Qing and Salahuddin, Sayeef and Muller, David A and Martin, Lane W and others},
  journal={Reviews of Modern Physics},
  volume={95},
  number={2},
  pages={025001},
  year={2023},
  publisher={APS}
}

@article{chen2021recent,
  title={Recent progress on topological structures in ferroic thin films and heterostructures},
  author={Chen, Shanquan and Yuan, Shuai and Hou, Zhipeng and Tang, Yunlong and Zhang, Jinping and Wang, Tao and Li, Kang and Zhao, Weiwei and Liu, Xingjun and Chen, Lang and others},
  journal={Advanced Materials},
  volume={33},
  number={6},
  pages={2000857},
  year={2021},
  publisher={Wiley Online Library}
}

@article{govinden2023spherical,
  title={Spherical ferroelectric solitons},
  author={Govinden, Vivasha and Prokhorenko, Sergei and Zhang, Qi and Rijal, Suyash and Nahas, Yousra and Bellaiche, Laurent and Valanoor, Nagarajan},
  journal={Nature Materials},
  volume={22},
  number={5},
  pages={553--561},
  year={2023},
  publisher={Nature Publishing Group UK London}
}

@article{nahas2015discovery,
  title={Discovery of stable skyrmionic state in ferroelectric nanocomposites},
  author={Nahas, Y and Prokhorenko, S and Louis, L and Gui, Z and Kornev, Igor and Bellaiche, Laurent},
  journal={Nature Communications},
  volume={6},
  number={1},
  pages={8542},
  year={2015},
  publisher={Nature Publishing Group UK London}
}

@article{zhang2017nanoscale,
  title={Nanoscale bubble domains and topological transitions in ultrathin ferroelectric films},
  author={Zhang, Qi and Xie, Lin and Liu, Guangqing and Prokhorenko, Sergei and Nahas, Yousra and Pan, Xiaoqing and Bellaiche, Laurent and Gruverman, Alexei and Valanoor, Nagarajan},
  journal={Advanced Materials},
  volume={29},
  number={46},
  pages={1702375},
  year={2017},
  publisher={Wiley Online Library}
}

@article{xu2025creating,
  title={Creating Ferroelectricity and Ultrahigh-Density Polar Skyrmion in Paraelectric Perovskite Oxide Monolayers by Moir{\'e} Engineering},
  author={Xu, Tao and Qian, Tao and Pang, Jiafei and Zhang, Jingtong and Li, Sheng and He, Ri and Wang, Jie and Shimada, Takahiro},
  journal={Research},
  volume={8},
  pages={0621},
  year={2025},
  publisher={AAAS}
}

@article{kern2022tailoring,
  title={Tailoring optical excitation to control magnetic skyrmion nucleation},
  author={Kern, L-M and Pfau, Bastian and Schneider, Michael and Gerlinger, Kathinka and Deinhart, Victor and Wittrock, Steffen and Sidiropoulos, Themistoklis and Engel, Dieter and Will, Ingo and G{\"u}nther, Christian M and others},
  journal={Physical Review B},
  volume={106},
  number={5},
  pages={054435},
  year={2022},
  publisher={APS}
}

@article{hu2022precise,
  title={Precise tuning of skyrmion density in a controllable manner by ion irradiation},
  author={Hu, Yue and Zhang, Senfu and Zhu, Yingmei and Song, Chengkun and Huang, Junfeng and Liu, Chen and Meng, Xuan and Deng, Xia and Zhu, Liu and Guan, Chaoshuai and others},
  journal={ACS Applied Materials \& Interfaces},
  volume={14},
  number={29},
  pages={34011--34019},
  year={2022},
  publisher={ACS Publications}
}

@article{schott2017skyrmion,
  title={The skyrmion switch: turning magnetic skyrmion bubbles on and off with an electric field},
  author={Schott, Marine and Bernand-Mantel, Anne and Ranno, Laurent and Pizzini, Stefania and Vogel, Jan and B{\'e}a, H{\'e}l{\`e}ne and Baraduc, Claire and Auffret, St{\'e}phane and Gaudin, Gilles and Givord, Dominique},
  journal={Nano Letters},
  volume={17},
  number={5},
  pages={3006--3012},
  year={2017},
  publisher={ACS Publications}
}

@article{zhong1994phase,
  title={Phase transitions in {{B}}a{{T}}i{{O}}\textsubscript{3} from first principles},
  author={Zhong, W and Vanderbilt, David and Rabe, KM},
  journal={Physical Review Letters},
  volume={73},
  number={13},
  pages={1861},
  year={1994},
  publisher={APS}
}

@article{zhong1995first,
  title={First-principles theory of ferroelectric phase transitions for perovskites: The case of {{B}}a{{T}}i{{O}}\textsubscript{3}},
  author={Zhong, W and Vanderbilt, David and Rabe, KM},
  journal={Physical Review B},
  volume={52},
  number={9},
  pages={6301},
  year={1995},
  publisher={APS}
}

@article{allen1992orbital,
  title={Orbital angular momentum of light and the transformation of Laguerre-Gaussian laser modes},
  author={Allen, Les and Beijersbergen, Marco W and Spreeuw, RJC and Woerdman, JP},
  journal={Physical review A},
  volume={45},
  number={11},
  pages={8185},
  year={1992},
  publisher={APS}
}

@article{xu2017designing,
  title={Designing lead-free antiferroelectrics for energy storage},
  author={Xu, Bin and {\'I}{\~n}iguez, Jorge and Bellaiche, Laurent},
  journal={Nature Communications},
  volume={8},
  number={1},
  pages={15682},
  year={2017},
  publisher={Nature Publishing Group UK London}
}

@article{jiang2018giant,
  title={Giant electrocaloric response in the prototypical {{P}}b({{M}}g, {{N}}b){{O}}\textsubscript{3} relaxor ferroelectric from atomistic simulations},
  author={Jiang, Zhijun and Nahas, Y and Prokhorenko, Sergei and Prosandeev, S and Wang, D and {\'I}{\~n}iguez, Jorge and Bellaiche, L},
  journal={Physical Review B},
  volume={97},
  number={10},
  pages={104110},
  year={2018},
  publisher={APS}
}

@article{ponomareva2005low,
  title={Low-dimensional ferroelectrics under different electrical and mechanical boundary conditions: Atomistic simulations},
  author={Ponomareva, I and Naumov, II and Bellaiche, L},
  journal={Physical Review B},
  volume={72},
  number={21},
  pages={214118},
  year={2005},
  publisher={APS}
}

@article{luk2020hopfions,
  title={Hopfions emerge in ferroelectrics},
  author={Luk’Yanchuk, I and Tikhonov, Y and Razumnaya, A and Vinokur, VM},
  journal={Nature Communications},
  volume={11},
  number={1},
  pages={2433},
  year={2020},
  publisher={Nature Publishing Group UK London}
}

@article{tikhonov2022polarization,
  title={Polarization topology at the nominally charged domain walls in uniaxial ferroelectrics},
  author={Tikhonov, Yurii and Maguire, Jesi R and McCluskey, Conor J and McConville, James PV and Kumar, Amit and Lu, Haidong and Meier, Dennis and Razumnaya, Anna and Gregg, John Martin and Gruverman, Alexei and others},
  journal={Advanced Materials},
  volume={34},
  number={45},
  pages={2203028},
  year={2022},
  publisher={Wiley Online Library}
}

@article{gao2025poincare,
  title={Poincar{\'e} sphere engineering of dynamical ferroelectric topological solitons},
  author={Gao, Lingyuan and Shen, Yijie and Prokhorenko, Sergei and Nahas, Yousra and Bellaiche, Laurent},
  journal={Physical Review B},
  volume={112},
  number={12},
  pages={L121102},
  year={2025},
  publisher={APS}
}

@article{xue2025observation,
  title={Observation of switchable polar skyrmion bubbles down to the atomic layers in van der Waals ferroelectric CuInP2S6},
  author={Xue, Fei and Zhang, Chenhui and Zheng, Sizheng and Tong, Peiran and Wang, Baoyu and Peng, Yong and Wang, Zhongyi and Xu, Haoran and He, Youshui and Zhou, Hongzhi and others},
  journal={Nature Communications},
  volume={16},
  number={1},
  pages={2349},
  year={2025},
  publisher={Nature Publishing Group UK London}
}

@article{han2022high,
  title={High-density switchable skyrmion-like polar nanodomains integrated on silicon},
  author={Han, Lu and Addiego, Christopher and Prokhorenko, Sergei and Wang, Meiyu and Fu, Hanyu and Nahas, Yousra and Yan, Xingxu and Cai, Songhua and Wei, Tianqi and Fang, Yanhan and others},
  journal={Nature},
  volume={603},
  number={7899},
  pages={63--67},
  year={2022},
  publisher={Nature Publishing Group UK London}
}

@article{jiang2025mechanically,
  title={Mechanically liberating polarization bubbles in van der Waals ferroelectrics},
  author={Jiang, Xingan and Wang, Tingjun and Zhang, Yixuan and Deng, Zunyi and Zhang, Xiangping and Zhu, Ruixue and Kang, Jiaqian and Yang, Xiangdong and Chen, Xue and Wang, Xiaolei and others},
  journal={Nature Materials},
  volume={24},
  number={12},
  pages={1942--1948},
  year={2025},
  publisher={Nature Publishing Group UK London}
}

@article{nagaosa2013topological,
  title={Topological properties and dynamics of magnetic skyrmions},
  author={Nagaosa, Naoto and Tokura, Yoshinori},
  journal={Nature Nanotechnology},
  volume={8},
  number={12},
  pages={899--911},
  year={2013},
  publisher={Nature Publishing Group UK London}
}

@article{wang2024giant,
  title={Giant electric field-induced second harmonic generation in polar skyrmions},
  author={Wang, Sixu and Li, Wei and Deng, Chenguang and Hong, Zijian and Gao, Han-Bin and Li, Xiaolong and Gu, Yueliang and Zheng, Qiang and Wu, Yongjun and Evans, Paul G and others},
  journal={Nature Communications},
  volume={15},
  number={1},
  pages={1374},
  year={2024},
  publisher={Nature Publishing Group UK London}
}

@article{li2025terahertz,
  title={Terahertz excitation of collective dynamics of polar skyrmions over a broad temperature range},
  author={Li, Wei and Wang, Sixu and Peng, Pai and Han, Haojie and Wang, Xinbo and Ma, Jing and Luo, Jianlin and Liu, Jun-Ming and Li, Jing-Feng and Nan, Ce-Wen and others},
  journal={Nature Physics},
  volume={21},
  number={12},
  pages={1965--1972},
  year={2025},
  publisher={Nature Publishing Group UK London}
}

@article{wang2025terahertz,
  title={Terahertz-field activation of polar skyrons},
  author={Wang, Huaiyu Hugo and Stoica, Vladimir A and Dai, Cheng and Pa{\'s}ciak, Marek and Das, Sujit and Yang, Tiannan and Gon{\c{c}}alves, Mauro AP and Kulda, Jiri and McCarter, Margaret R and Mangu, Anudeep and others},
  journal={Nature Communications},
  volume={16},
  number={1},
  pages={8994},
  year={2025},
  publisher={Nature Publishing Group UK London}
}

@article{fert2013skyrmions,
  title={Skyrmions on the track},
  author={Fert, Albert and Cros, Vincent and Sampaio, Joao},
  journal={Nature Nanotechnology},
  volume={8},
  number={3},
  pages={152--156},
  year={2013},
  publisher={Nature Publishing Group UK London}
}

@article{tomasello2014strategy,
  title={A strategy for the design of skyrmion racetrack memories},
  author={Tomasello, Riccardo and Martinez, E and Zivieri, Roberto and Torres, Luis and Carpentieri, Mario and Finocchio, Giovanni},
  journal={Scientific Reports},
  volume={4},
  number={1},
  pages={6784},
  year={2014},
  publisher={Nature Publishing Group UK London}
}

@article{luo2021skyrmion,
  title={Skyrmion devices for memory and logic applications},
  author={Luo, Shijiang and You, Long},
  journal={APL Materials},
  volume={9},
  number={5},
  pages={050901},
  year={2021},
  publisher={AIP Publishing}
}

@article{pereira2019theoretical,
  title={Theoretical guidelines to create and tune electric skyrmion bubbles},
  author={Pereira Gon{\c{c}}alves, Mauro Ant{\'o}nio and Escorihuela-Sayalero, Carlos and Garca-Fern{\'a}ndez, Pablo and Junquera, Javier and {\'I}{\~n}iguez, Jorge},
  journal={Science Advances},
  volume={5},
  number={2},
  pages={eaau7023},
  year={2019},
  publisher={American Association for the Advancement of Science}
}

@article{ang2020electrical,
  title={Electrical control of skyrmion density via skyrmion-stripe transformation},
  author={Ang, Calvin Ching Ian and Gan, Weiliang and Wong, Grayson Dao Hwee and Lew, Wen Siang},
  journal={Physical Review Applied},
  volume={14},
  number={5},
  pages={054048},
  year={2020},
  publisher={APS}
}

@article{wang2018ferroelectrically,
  title={Ferroelectrically tunable magnetic skyrmions in ultrathin oxide heterostructures},
  author={Wang, Lingfei and Feng, Qiyuan and Kim, Yoonkoo and Kim, Rokyeon and Lee, Ki Hoon and Pollard, Shawn D and Shin, Yeong Jae and Zhou, Haibiao and Peng, Wei and Lee, Daesu and others},
  journal={Nature Materials},
  volume={17},
  number={12},
  pages={1087--1094},
  year={2018},
  publisher={Nature Publishing Group UK London}
}

@article{denker2023size,
  title={Size and density control of skyrmions with picometer CoFeB thickness variations—observation of zero-field skyrmions and skyrmion merging},
  author={Denker, Christian and Nielsen, S{\"o}ren and Lage, Enno and R{\"o}mer-Stumm, Malte and Heyen, Hauke and Junk, Yannik and Walowski, Jakob and Waldorf, Konrad and M{\"u}nzenberg, Markus and McCord, Jeffrey},
  journal={Journal of Physics D: Applied Physics},
  volume={56},
  number={49},
  pages={495302},
  year={2023},
  publisher={IOP Publishing}
}

@article{lone2024controlling,
  title={Controlling the skyrmion density and size for quantized convolutional neural network},
  author={Lone, Aijaz H and Ganguly, Arnab and Li, Hanrui and El-Atab, Nazek and Setti, Gianluca and Das, Gobind and Fariborzi, Hossein},
  journal={IEEE Access},
  year={2024},
  publisher={IEEE}
}

@article{liefferink2025effective,
  title={Effective Theory of Ultrafast Skyrmion Nucleation},
  author={Liefferink, Rein and K{\"o}rber, Lukas and Gerlinger, Kathinka and Pfau, Bastian and B{\"u}ttner, Felix and Mentink, Johan H},
  journal={arXiv preprint arXiv:2504.11013},
  year={2025}
}

@article{prinz2023orbital,
  title={Orbital angular momentum in nanoplasmonic vortices},
  author={Prinz, Eva and Hartelt, Michael and Spektor, Grisha and Orenstein, Meir and Aeschlimann, Martin},
  journal={ACS Photonics},
  volume={10},
  number={2},
  pages={340--367},
  year={2023},
  publisher={ACS Publications}
}

@article{koya2023advances,
  title={Advances in ultrafast plasmonics},
  author={Koya, Alemayehu Nana and Romanelli, Marco and Kuttruff, Joel and Henriksson, Nils and Stefancu, Andrei and Grinblat, Gustavo and De Andres, Aitor and Schnur, Fritz and Vanzan, Mirko and Marsili, Margherita and others},
  journal={Applied Physics Reviews},
  volume={10},
  number={2},
  pages={021318},
  year={2023},
  publisher={AIP Publishing}
}

@article{wang2023entangled,
  title={Entangled polarizations in ferroelectrics: A focused review of polar topologies},
  author={Wang, YJ and Tang, YL and Zhu, YL and Ma, XL},
  journal={Acta Materialia},
  volume={243},
  pages={118485},
  year={2023},
  publisher={Elsevier}
}

@article{makeev2024quasihexagonal,
  title={Quasihexagonal arrays of electric-skyrmion bubbles in thin-film ferroelectrics: Pattern formation and structure},
  author={Makeev, Maxim A and Rijal, Suyash and Prokhorenko, Sergei and Nahas, Yousra and Bellaiche, Laurent},
  journal={Physical Review B},
  volume={110},
  number={14},
  pages={144113},
  year={2024},
  publisher={APS}
}

@article{garcia1998electromechanical,
  title={Electromechanical behavior of {BaTiO$_3$} from first principles},
  author={Garcia, Alberto and Vanderbilt, David},
  journal={Applied Physics Letters},
  volume={72},
  number={23},
  pages={2981--2983},
  year={1998},
  publisher={American Institute of Physics}
}

@article{bhattacharya2020creation,
  title={Creation and annihilation of non-volatile fixed magnetic skyrmions using voltage control of magnetic anisotropy},
  author={Bhattacharya, Dhritiman and Razavi, Seyed Armin and Wu, Hao and Dai, Bingqian and Wang, Kang L and Atulasimha, Jayasimha},
  journal={Nature Electronics},
  volume={3},
  number={9},
  pages={539--545},
  year={2020},
  publisher={Nature Publishing Group UK London}
}

@article{chen2024all,
  title={All-electrical skyrmionic magnetic tunnel junction},
  author={Chen, Shaohai and Lourembam, James and Ho, Pin and Toh, Alexander KJ and Huang, Jifei and Chen, Xiaoye and Tan, Hang Khume and Yap, Sherry LK and Lim, Royston JJ and Tan, Hui Ru and others},
  journal={Nature},
  volume={627},
  number={8004},
  pages={522--527},
  year={2024},
  publisher={Nature Publishing Group UK London}
}

@article{je2018creation,
  title={Creation of magnetic skyrmion bubble lattices by ultrafast laser in ultrathin films},
  author={Je, Soong-Geun and Vallobra, Pierre and Srivastava, Titiksha and Rojas-S{\'a}nchez, Juan-Carlos and Pham, Thai Ha and Hehn, Michel and Malinowski, Gregory and Baraduc, Claire and Auffret, St{\'e}phane and Gaudin, Gilles and others},
  journal={Nano Letters},
  volume={18},
  number={11},
  pages={7362--7371},
  year={2018},
  publisher={ACS Publications}
}

@article{berg1981definition,
  title={Definition and statistical distributions of a topological number in the lattice O (3) $\sigma$-model},
  author={Berg, B and L{\"u}scher, Martin},
  journal={Nuclear Physics B},
  volume={190},
  number={2},
  pages={412--424},
  year={1981},
  publisher={Elsevier}
}

@article{wu2013graphene,
  title={Graphene/liquid crystal based terahertz phase shifters},
  author={Wu, Yang and Ruan, Xuezhong and Chen, Chih-Hsin and Shin, Young Jun and Lee, Youngbin and Niu, Jing and Liu, Jingbo and Chen, Yuanfu and Yang, Kun-Lin and Zhang, Xinhai and others},
  journal={Optics express},
  volume={21},
  number={18},
  pages={21395--21402},
  year={2013},
  publisher={Optical Society of America}
}

@article{fischer2023simultaneous,
  title={Simultaneous direct measurement of the electrocaloric and dielectric dynamics of ferroelectrics with microsecond temporal resolution},
  author={Fischer, Jonas and D{\"o}ntgen, Jago and Molin, Christian and Gebhardt, SE and Hambal, Yusra and Shvartsman, VV and Lupascu, DC and H{\"a}gele, Daniel and Rudolph, J{\"o}rg},
  journal={Review of Scientific Instruments},
  volume={94},
  number={4},
  pages={043906},
  year={2023},
  publisher={AIP Publishing}
}

@article{wang2015broadband,
  title={Broadband tunable liquid crystal terahertz waveplates driven with porous graphene electrodes},
  author={Wang, Lei and Lin, Xiao-Wen and Hu, Wei and Shao, Guang-Hao and Chen, Peng and Liang, Lan-Ju and Jin, Biao-Bing and Wu, Pei-Heng and Qian, Hao and Lu, Yi-Nong and others},
  journal={Light: Science \& Applications},
  volume={4},
  number={2},
  pages={e253--e253},
  year={2015},
  publisher={Nature Publishing Group}
}

@article{de2026enhanced,
  title={Enhanced Terahertz Photoresponse via Acoustic Plasmon Cavity Resonances in Scalable Graphene},
  author={De Fazio, Domenico and Castilla, Sebasti{\'a}n and Soundarapandian, Karuppasamy P and Slipchenko, Tetiana and Vangelidis, Ioannis and Marconi, Simone and Bertini, Riccardo and Petrica, Vlad and Hao, Yang and Principi, Alessandro and others},
  journal={arXiv preprint arXiv:2601.16604},
  year={2026}
}

@article{ko2019improvement,
  title={Improvement of reliability and dielectric breakdown strength of Nb-doped lead zirconate titanate films via microstructure control of seed},
  author={Ko, Song Won and Zhu, Wanlin and Fragkiadakis, Charalampos and Borman, Trent and Wang, Ke and Mardilovich, Peter and Trolier-McKinstry, Susan},
  journal={Journal of the American Ceramic Society},
  volume={102},
  number={3},
  pages={1211--1217},
  year={2019},
  publisher={Wiley Online Library}
}

@article{nguyen2019experimental,
  title={Experimental evidence of breakdown strength and its effect on energy-storage performance in normal and relaxor ferroelectric films},
  author={Nguyen, Minh D and Nguyen, Chi TQ and Vu, Hung N and Rijnders, Guus},
  journal={Current Applied Physics},
  volume={19},
  number={9},
  pages={1040--1045},
  year={2019},
  publisher={Elsevier}
}

@article{kwak2011dielectric,
  title={Dielectric characteristics of Pb (Zr, Ti) O3 films on MgO single crystal substrate by terahertz time domain spectroscopy},
  author={Kwak, Min Hwan and Kang, Seung Beom and Kim, Ki-Chul and Jeong, Se Young and Kim, Sungil and Yoo, Byung Hwa and Chung, Dong Chul and Ryu, Han Cheol and Jun, Dong Suk and Paek, Mun Cheol and others},
  journal={Ferroelectrics},
  volume={422},
  number={1},
  pages={19--22},
  year={2011},
  publisher={Taylor \& Francis}
}

@inproceedings{mcpherson2002proposed,
  title={Proposed universal relationship between dielectric breakdown and dielectric constant},
  author={McPherson, J and Kim, J and Shanware, A and Mogul, H and Rodriguez, J},
  booktitle={Digest. International Electron Devices Meeting,},
  pages={633--636},
  year={2002},
  organization={IEEE}
}
\clearpage

\begin{figure}
\includegraphics[width = 160mm]{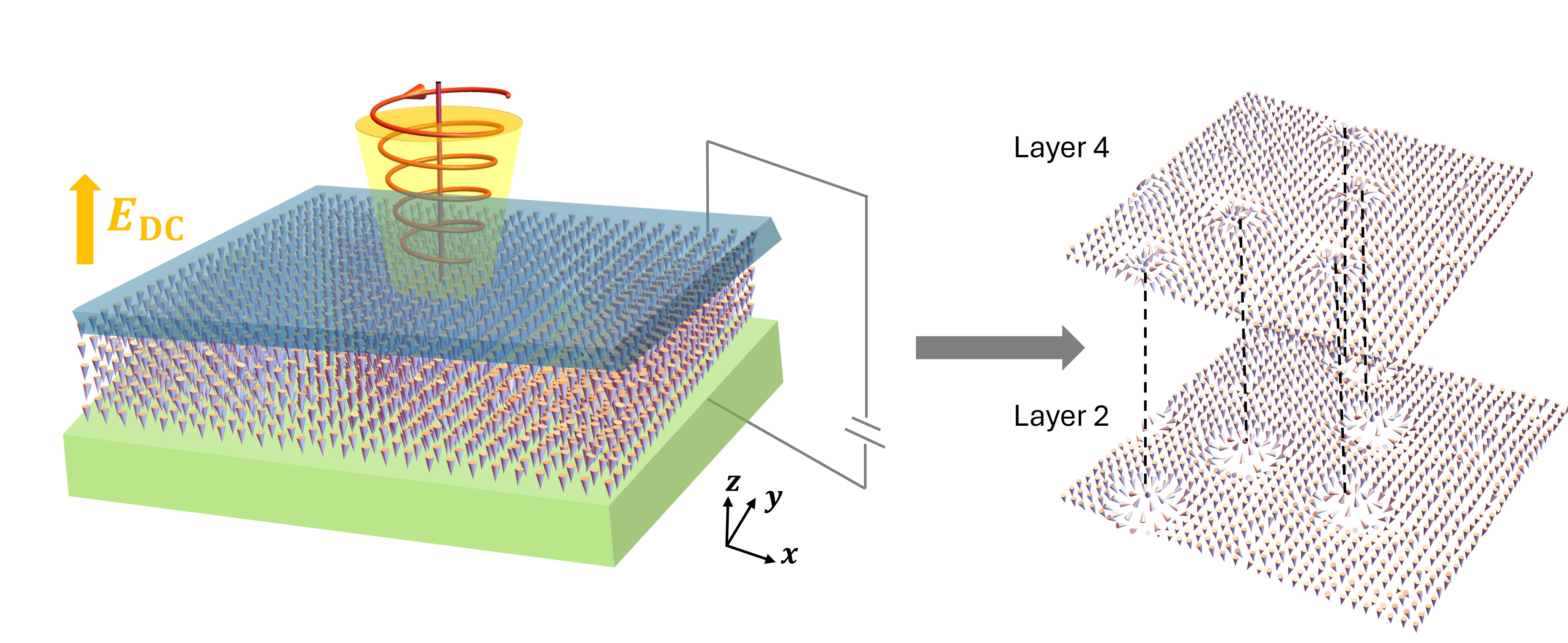}
\caption{\textbf{Schematic plot of twisted light interacting with a ferroelectric PZT ultrathin film}. The initial dipole configuration is a $80\times80\times5$ down-poled monodomain. Electrodes and substrates denoted by the blue and green pads are attached to the top and the bottom of the film respectively, and they are connected to an external voltage to apply a DC electric field along the out-of-plane $z$ direction. The field of twisted light can be written as  $\vec{E}(\vec{r},t) = E_0(\frac{\sqrt{2}r}{w})e^{-\frac{{r}^2}{w^2}} \big{(}\cos{( \phi + \omega t)} \vec{e}_x + \sin{( \phi + \omega t)} \vec{e}_y \big{)}$, where $w$, $\vec{r}$ and $\phi$ denote the beam radius, radial distance and azimuthal angle on the plane respectively. On the right, emergent polar skyrmion bubbles on two subsurface layers are shown. These bubbles, as shown below, are three-dimensional (3D) flux-closure structures, and they share the same in-plane positions between two layers but with opposite helicities and appear as divergent on layer 4 and convergent on layer 2.}
\end{figure}

\begin{figure}
\includegraphics[width = 160mm]{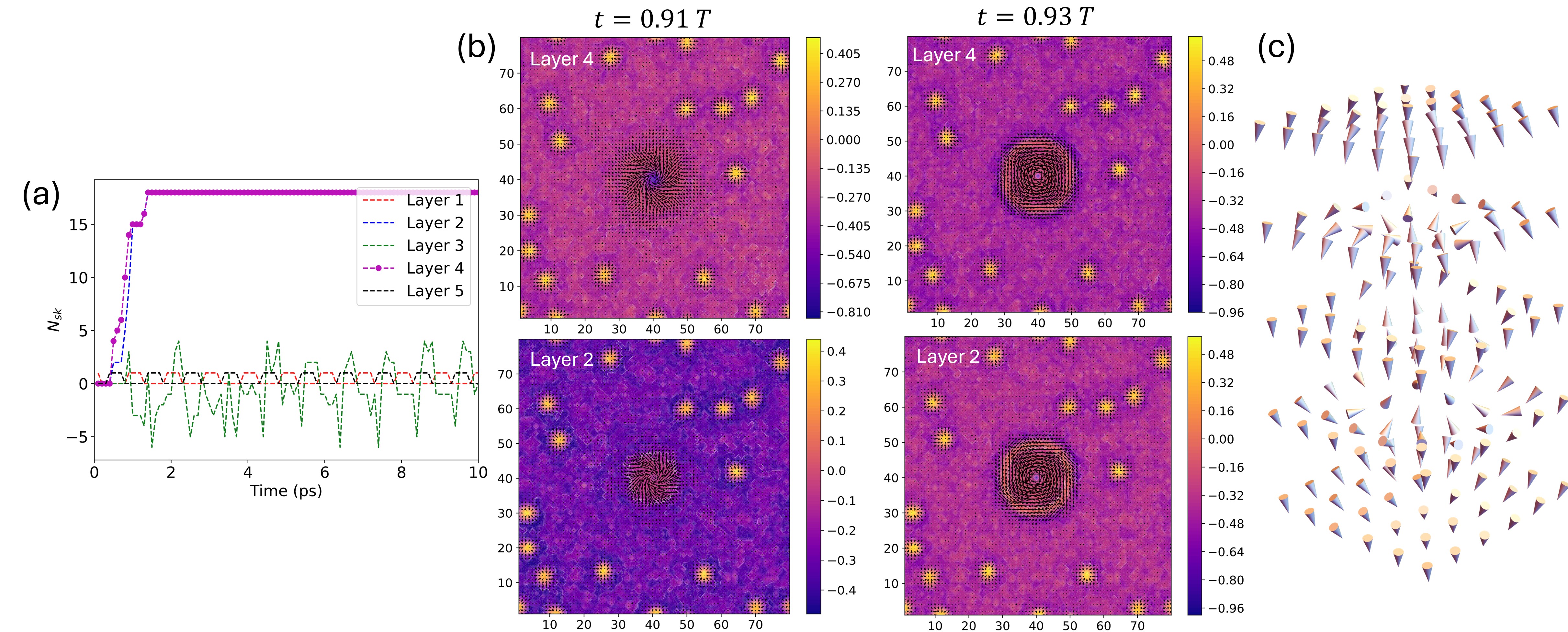}
\caption{\textbf{Skyrmion bubble formation on subsurface layers.} (a) Variation of $N_{\rm{sk}}$ on each layer with time when $E_{0} = 180$ MV/cm and $E_{\rm{DC}} = 0.725$ MV/cm. Layer 1 and 5 denote the bottom and top surface layers, and layer 2 and 4 denote the subsuface layers adjacent to the bottom and the top surface. The period of the field $T$ is 1 ps. (b) In-plane views of dipolar configurations on two subsurface layers at different times $t = 0.91T$ and $t = 0.93T$. $x$ and $y$ axes denote the unit cell number along the respective direction, and the length/width of each cell is $\sim$0.4 nm. The relative out-of-plane polarization magnitude $p_{z,i}(t)/p_{z,\rm{max}}(t)$ is denoted by the colorbar, where $p_{z,\rm{max}}(t)$ denotes the maximum $p_z$ throughout the simulation cell at time $t$, and that corresponds to a polarization value of $1.2\sim1.5 \ C/m^2$. The in-plane $p_{x,y}$ components are denoted by the arrow. (c) 3D visualization of a flux-closure skyrmion bubble, where the bottom part (layer 2) and top part (layer 4)  exhibit opposite helicity for $p_{x,y}$, and the middle layer (layer 3) has a core with vanishing $p_{x,y}$ and  opposite/upward $p_z$.}
\end{figure}

\begin{figure}
\includegraphics[width = 160mm]{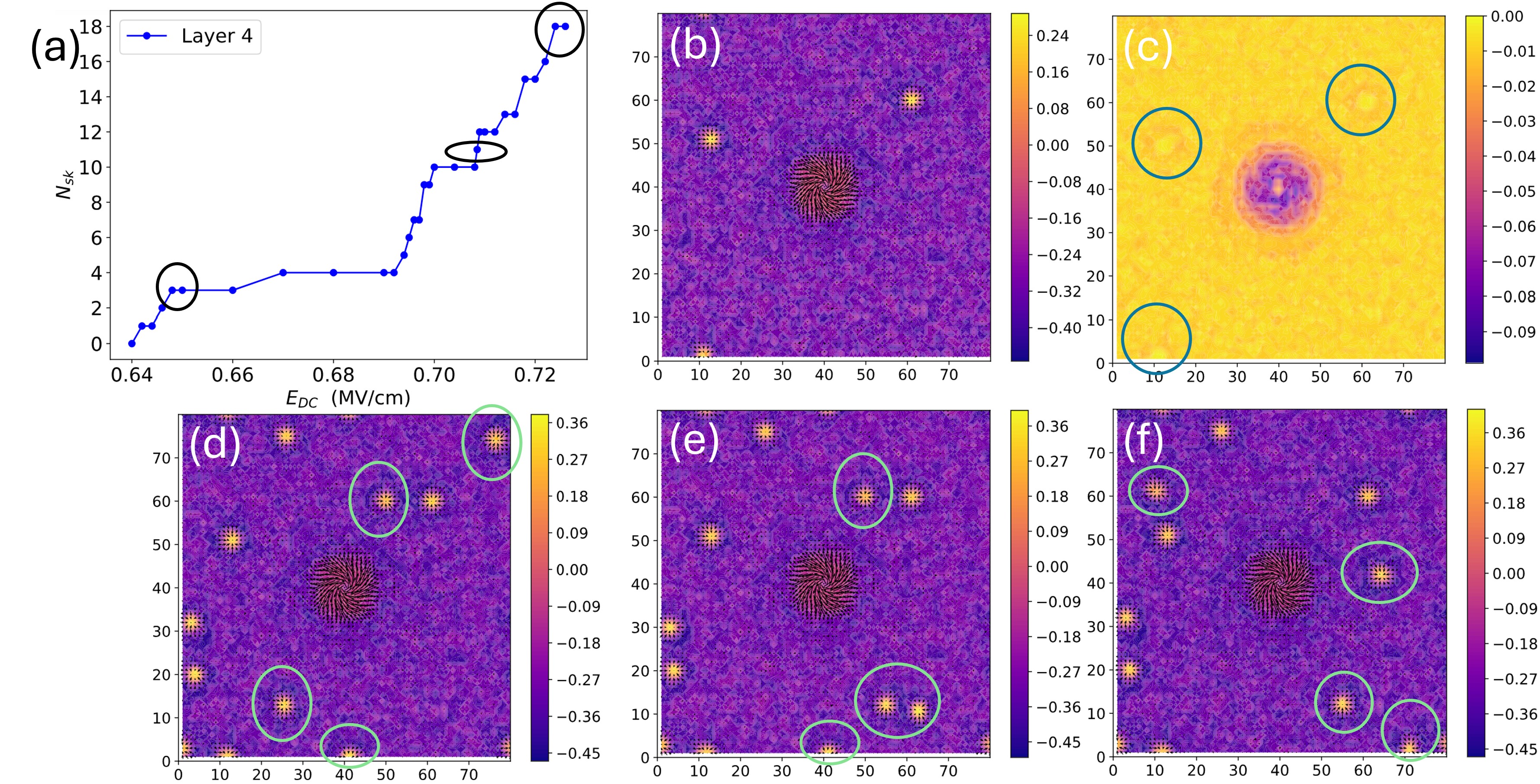}
\caption{ \textbf{Skyrmion bubble density modulatted by DC electric field magnitude.} (a) Variation of $N_{\rm{sk}}$ on subsurface layers with $E_{\rm{DC}}$ when $E_{0} = 180$ MV/cm. (b) The in-plane view of the dipole configuration on layer 4 when there are three bubbles at $t = 0.91T$ for $E_{\rm{DC}}= 0.65 $ MV/cm. These bubbles are at identical locations with different initial thermalizations. (c) The corresponding spatial map of dipole-dipole interaction energy for three bubbles; locations where bubbles reside have slightly lower energies and are highlighted by circles. (d)-(f) Similar to (b), but for eleven bubbles when $E_{\rm{DC}} \approx 0.71$ MV/cm. Three panels correspond to simulations with different initial thermalizations, and four out of eleven bubbles that do not appear at the same locations across simulations are highlighted by circles.}
\end{figure}

\begin{figure} 
\includegraphics[width = 160mm]{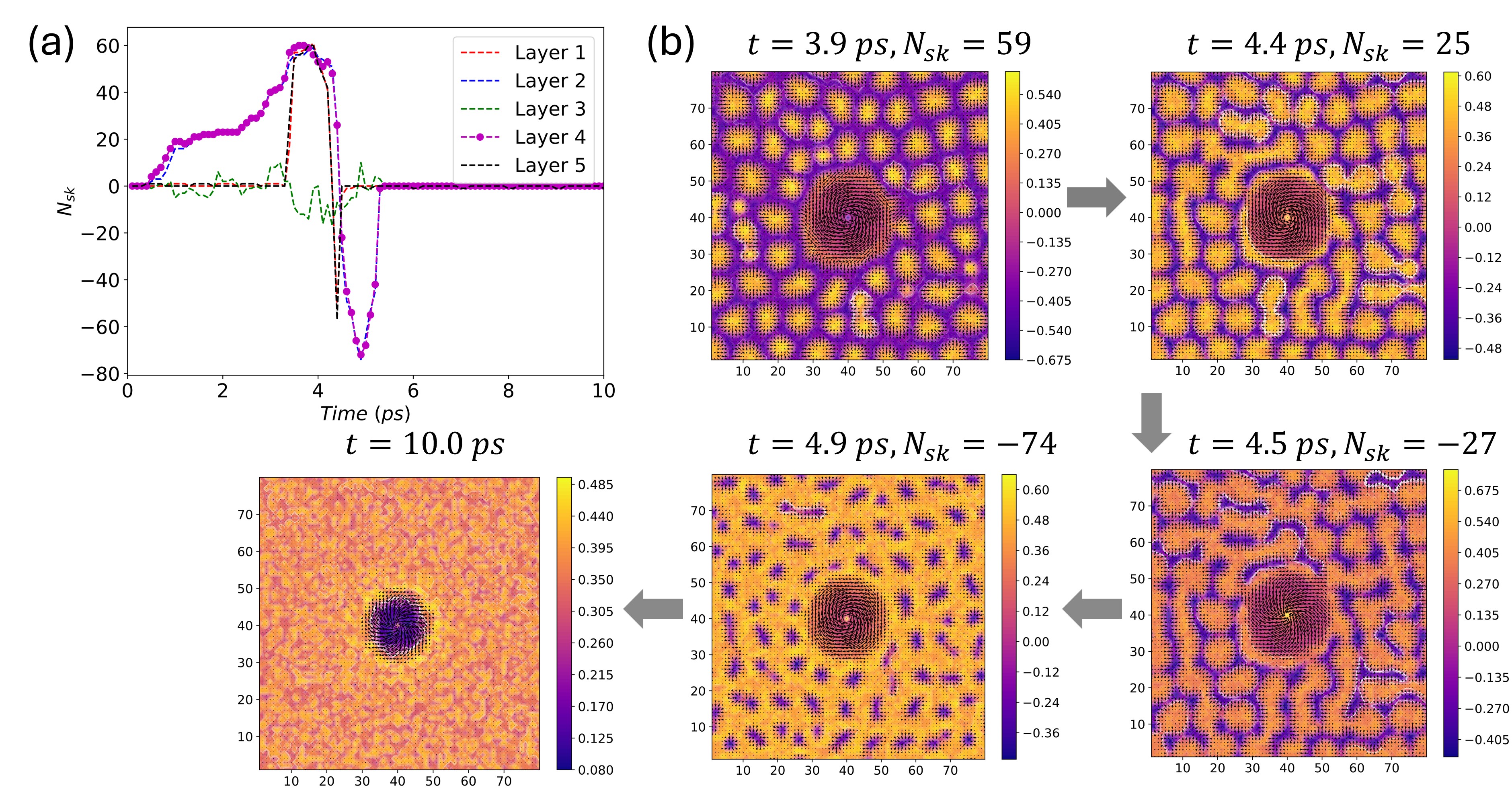}
\caption{\textbf{Annihilation of polar skyrmion bubbles}. (a) Variation of $N_{\rm{sk}}$ on each layer with time when $E_{0} = 180$ MV/cm and $E_{\rm{DC}} = 0.73$ MV/cm. (b) In-plane views of dipole configuration on subsurface layer 4 at different stages during the transition, as described in details in the text. }
\end{figure}

\begin{figure}
\includegraphics[width = 120mm]{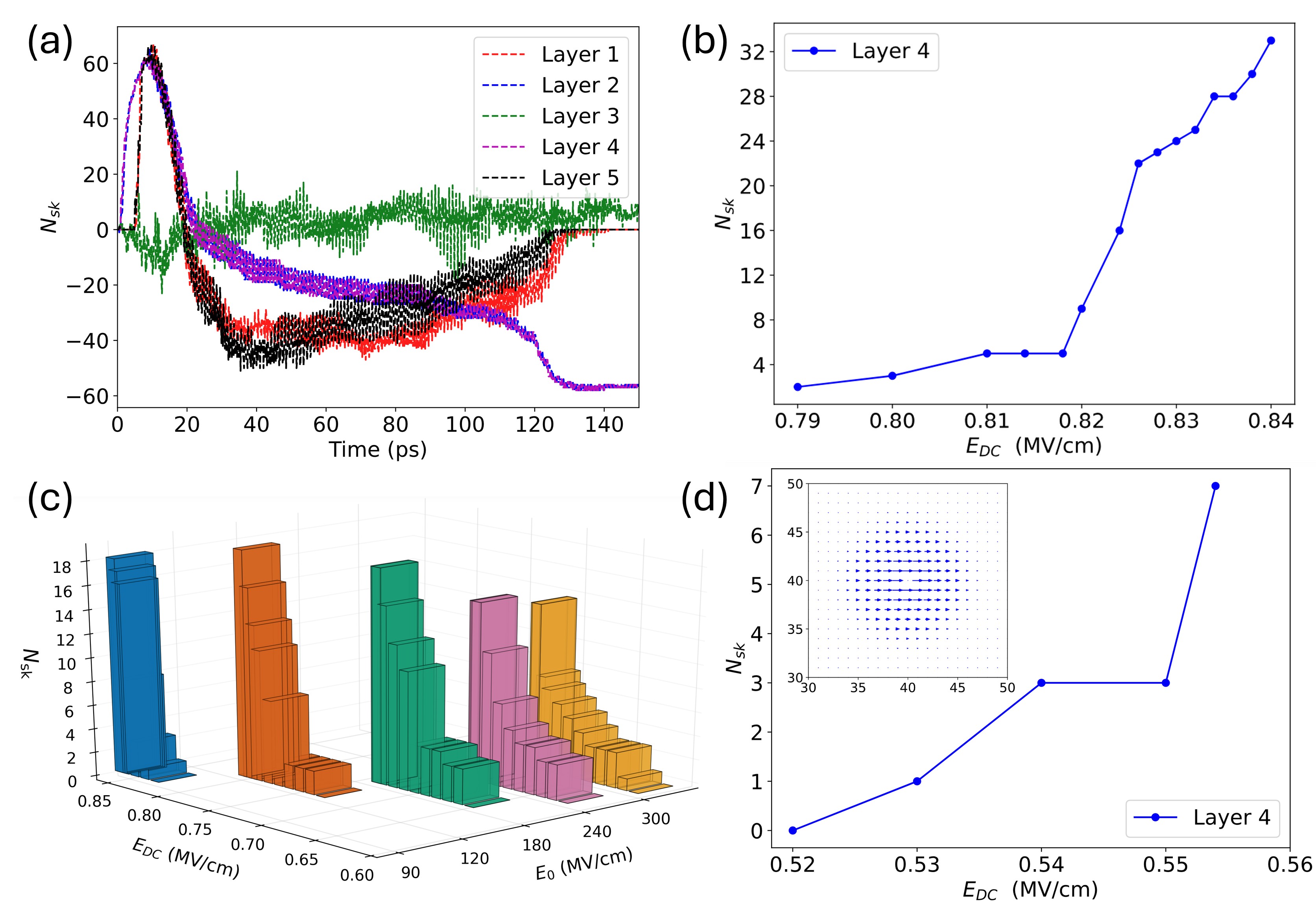}
\caption{\textbf{Tunability of skyrmion bubble density.} (a) Variation of $N_{\rm{sk}}$ with time up to 150 ps on each layer when twisted light has a larger beam radius $w = 10$ u.c., with $E{_0} = 80$ MV/cm, $E_{\rm{DC}} = 0.8$ MV/cm. (b) Variation of $N_{\rm{sk}}$ with $E_{\rm{DC}}$ on each layer in a  $100 \times 100 \times 5$ simulation cell, with $E{0} = 180$ MV/cm, $w = 5$ u.c.. (c) Histogram showing the dependence of $N_{\rm{sk}}$ on both $E_{\rm{0}}$ and $E_{\rm{DC}}$. (d) Variation of $N_{\rm{sk}}$ with $E_{\rm{DC}}$ for a circularly-polarized Gaussian pulse, whose field is described as $\vec{E}(\vec{r},t) = E_0e^{-\frac{{r}^2}{w^2}} \big{(}\cos{( \omega t)} \vec{e}_x + \sin{(\omega t)} \vec{e}_y \big{)}$, with $w =$ 5 u.c. and  $E_0= 180$MV/cm. The field profile at integer multiples of $T$ is illustrated in the inset.  }
\end{figure}

\clearpage
\foreach \x in {1,...,6}
{

 \includepdf[pages={\x}]{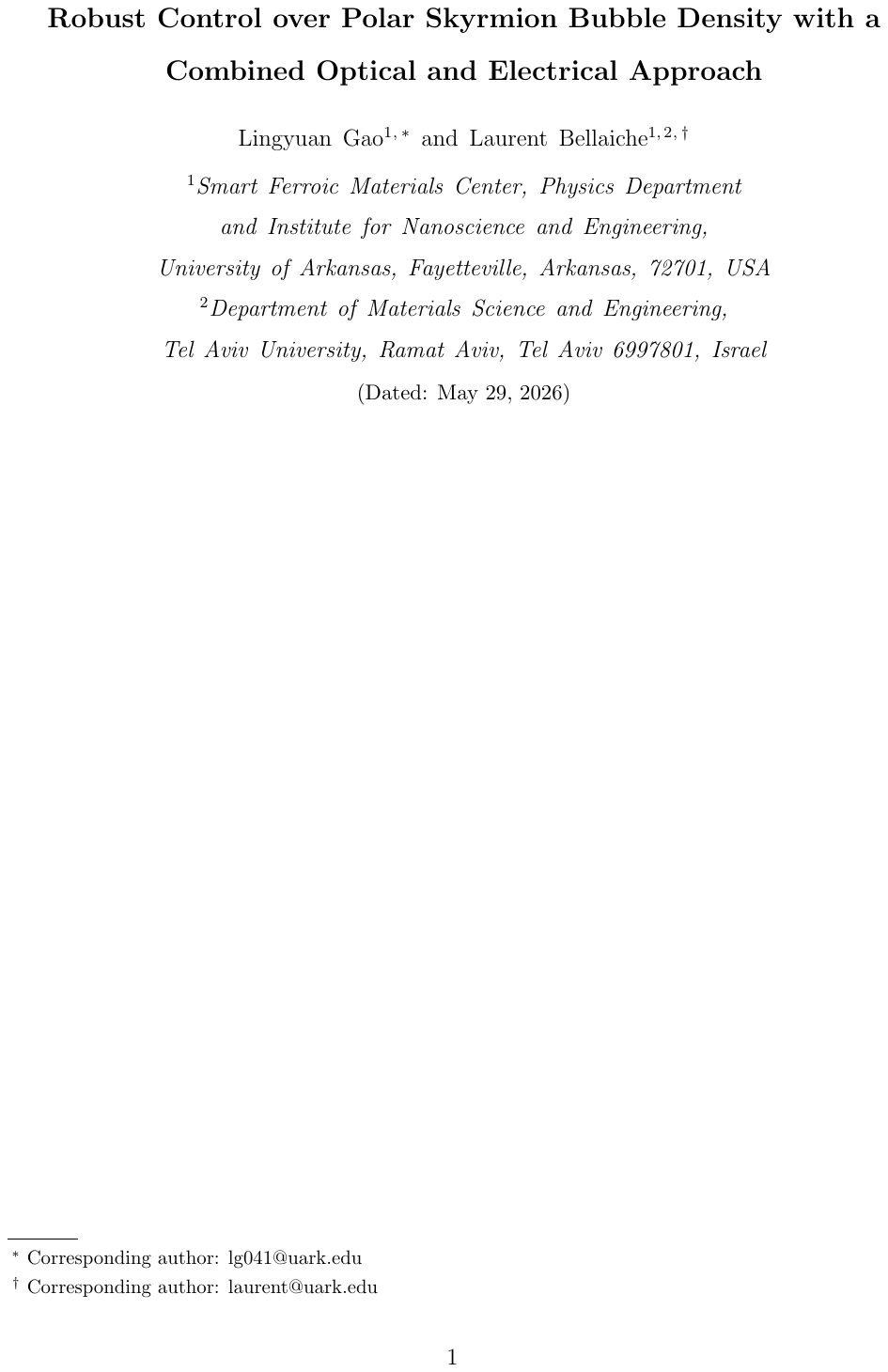}
}

\end{document}